\documentclass{article}    
\usepackage{setspace}

\usepackage{graphicx}
\usepackage{dcolumn}
\usepackage{bm}

\begin{document}           

\title{Testing Cosmological General Relativity against high redshift observations}

\author{\textbf{John G. Hartnett}\\
School of Physics, the University of Western Australia,\\
 35 Stirling Hwy, Crawley 6009 WA Australia\\
\textit{john@physics.uwa.edu.au}\\
\textbf{Firmin J. Oliveira}\\
Joint Astronomy Centre Hilo, Hawai`i 96720 \\
\textit{firmin@jach.hawaii.edu}}

\date{ }
\maketitle
\begin{abstract}

  Several key relations are derived for Cosmological General Relativity
  which are used in standard observational cosmology. These include the luminosity distance,
  angular size, surface brightness and matter density.  These relations are used to fit
  type Ia supernova (SNe Ia) data, giving consistent, well behaved fits over a broad
  range of redshift $0.1 < z < 2$.  The best fit to the data for the local density parameter
  is $\Omega_{m} = 0.0401 \pm 0.0199$. Because $\Omega_{m}$ is within the baryonic budget there
  is no need for any dark matter to account for the SNe Ia redshift luminosity data.
  From this local density it is determined that the redshift where the universe expansion
  transitions from deceleration to acceleration is $z_{t}= 1.095\, {}^{+0.264}_{-0.155}$. 
  Because the fitted data covers the range of the predicted transition redshift $z_{t}$,
  there is no need for any dark energy to account for the expansion rate transition.
  We conclude that the expansion is now accelerating and that the transition from a closed
  to an open universe occurred about $8.54\, {\rm Gyr}$ ago.

\end{abstract}

\section{\bf{INTRODUCTION}}

Carmeli's cosmology, also referred to as Cosmological General Relativity (CGR), is a
  space-velocity theory of the expanding universe. It is a description of the universe at a
  particular fixed epoch of cosmic time $t$. In CGR time is measured from the present back
  toward the beginning. The theory assumes the Hubble law as fundamental.  The observables are the coordinates of Hubble; proper distance and velocity of the expansion of the Universe. In practice, not velocity but redshift is used.  CGR incorporates this basic law into a general $4D$ Riemannian geometrical theory  satisfying the Einstein field equations (Ref. \cite{Carmeli2002}, appendix A).
  
In order to compare the theoretical predicted redshift distance modulus relation of CGR with the distance modulii derived from type Ia supernova data, firstly, luminosity distance must be determined in this theory. Secondly, we need to model correctly the variation of matter density with redshift in the Universe. In the following, we determine a few key relations that are used in the subsequent analysis. Then we compare the theoretical distance modulii with those measured, resulting in a good fit without the need to assume the existence of dark energy or dark matter.

\section{\bf{LUMINOSITY DISTANCE}}

  Suppose $L$ is the total energy emitted per unit time by a source galaxy at the epoch $t$ (that is, in the rest frame of the galaxy) to be received by an observer at the present time $t=0$. Therefore we can write 
  \begin{equation} \label{eq:lumin}
     dL = L I(\lambda) d\lambda,
  \end{equation}
where $I$ is its (normalized) intensity distribution -- a function of wavelength $\lambda$. In CGR, times at cosmological distances add according to a relativistic addition law \cite{Carmeli-1A} when referred to the observer at $t = 0$. Hence instead of the time interval $\Delta t$, we get 
\begin{equation} \label{eq:deltat}
     \Delta t \rightarrow \frac{t+ \Delta t}{1+\frac{t \Delta t}{\tau^2}} - t
     = \Delta t \left\{1-\frac{t^2}{\tau^2}\right\},
  \end{equation}
where $\tau \approx H_0^{-1}$ is the Hubble-Carmeli time constant.  From this it can be shown \cite{Hartnett2007} that the luminosity $L_0$ of a source at the present time is related to the luminosity $L$ of an identical source which emitted at time $t$ by
   \begin{equation} \label{eq:luminosity}
L_0 =  L \left\{1-\frac{t^2}{\tau^2}\right\}.
  \end{equation}
For the source at distance $r$, redshift $z$, emission wavelength $\lambda_0 / (1+z)$ and the luminosity (\ref{eq:luminosity}), it is straight forward to show \cite{Narlikar2002} that the observed flux integrated over all wavelengths is 
  \begin{equation} \label{eq:bol}
   \mathcal{F}_{bol} =  \frac{L_{bol}}{(1+z)^2}  \left\{1-\frac{t^2}{\tau^2}\right\}
            \frac{1}{4\pi r^2} = \frac{L_{bol}}{4 \pi \mathcal{D}_L^2},
  \end{equation} 
  where $L_{bol}$ is the absolute bolometric luminosity of the source galaxy. Therefore the luminosity distance $\mathcal{D}_L$
 in CGR is expressed as
    \begin{equation} \label{eq:lumdistance}
 \mathcal{D}_L =  r(1+z)\left\{1-\frac{t^2}{\tau^2}\right\}^{-1/2}.
  \end{equation}

 It is clear that this expression for the luminosity distance in CGR when compared to that in the FRW theory has the extra factor $(1-t^2/\tau^2)^{-1/2}$. Hence we expect the luminosity distance to be greater in CGR than in FRW theory.
\\ 
\section{\label{sec:Angular}\bf{ANGULAR SIZE}}  

The line element in CGR \cite{Carmeli2002b}
\begin{equation} \label{eqn:lineelement}
ds^2= \tau^{2}dv^{2}-\left(1 + (1-\Omega)\frac {r^{2}}{c^{2}\tau^{2}}\right)^{-1}dr^{2}-r^2(d\theta^2+ sin^2\theta d\phi^2),
\end{equation}
represents a spherically symmetric isotropic universe. See Ref \cite{Carmeli2002, Carmeli2002b} for details. The expansion is observed 
at a definite time and therefore $dt = 0$ and hence doesn't appear in (\ref{eqn:lineelement}). Carmeli solved (\ref{eqn:lineelement}) 
with the null condition $ds = 0$ and isotropy ($d\theta=d\phi=0$) from which it follows that the proper distance $r$ in spherically
 symmetric coordinates can be written as
\begin{equation} \label{eqn:rctau}
\frac {r} {c \tau}= \frac {\sinh \left(\beta \sqrt{1-\Omega}\right)} {\sqrt{1-\Omega}} ,  
\end{equation}
where $\beta =t/\tau =v/c$ and $\Omega$ is matter density, a function of redshift $z$. Also $\beta$ can be written as a function of
 redshift 
\begin{equation} \label{eqn:defbeta}
\beta = \frac{(1+z)^2-1}{(1+z)^2+1}.
\end{equation}

Now in CGR there is no scale factor like in the FRW theory but we can similarly define an expansion factor as $(1+z)^{-1}$. If we then
 make the substitution for the matter density $\Omega = \Omega_m (1+z)^{3}$, where $\Omega_m$ is the matter density at the current epoch,
 the proper distance (\ref{eqn:rctau}) can be rewritten as a function of $(1+z)$,
\begin{equation} \label{eqn:rctau2}
r = c \tau  \sinh \left(\beta \sqrt{1-\Omega_m (1+z)^{3}}\right)/\sqrt{1-\Omega_m (1+z)^{3}}.  
\end{equation}
 
For a proper comparison with FRW theory we must use the FRW equivalent of $r/(1+z)$, which is the Hubble distance $D_1$ when the light
 we observe left the galaxy at redshift $z$  and is given by
\begin{equation} \label{eqn:rFRW}
  D_1 = \frac {2cH^{-1}_0} {(1+z)}\left \{1-\frac{1}{\sqrt{1+z}}\right \},  
\end{equation}
where a deceleration parameter $q_0 = 1/2$ has been used. The angular size of the source galaxy in FRW theory is 
\begin{equation} \label{eqn:dtheta}
  \Delta {\theta} = \frac {d} {D_1},  
\end{equation}
where $d$ is the actual diameter of the source galaxy and the angular distance $D_1$ is taken from (\ref{eqn:rFRW}). 

In CGR the angular distance $\mathcal{D}_A$ is defined identically with (\ref{eqn:dtheta}) 
\begin{equation} \label{eqn:dthetaCGR}
  \Delta {\theta} = \frac {d} {\mathcal{D}_A}  \, ,
\end{equation}
where the functional form for $\mathcal{D}_A$ is determined by its relationship to the luminosity distance $\mathcal{D}_{L}$. To show
 how $\mathcal{D}_L$ and $\mathcal{D}_A$ are related we look at the flux $F_{\theta}$ from a distant source of extent $d$ which subtends
 an angle $\Delta{\theta}$ on the sky \cite{nedwright}
\begin{equation}
  F_{\theta} = \Delta{\theta}^2 \sigma \, T^4_{o} \, ,  \label{eqn:FthetaCGR}  
\end{equation}
where $\sigma$ is the Stephan-Boltzmann constant and $T_{o}$ is the observed temperature of the source.  Equating fluxes from 
(\ref{eq:bol}) and (\ref{eqn:FthetaCGR}), substituting for $\Delta{\theta}$ from (\ref{eqn:dthetaCGR}) and substituting
 $L_{bol} = 4 \pi d^2 \sigma T^4_{e}$  with $T_{e}$ the source temperature we get
\begin{equation}
   \frac{T^4_{e}} {\mathcal{D}^2_{L}} = \frac{T^4_{o}} {\mathcal{D}^2_{A}} \, . \label{eq:TobsDA_TemDL}
\end{equation}
Since for a blackbody at temperature $T$ the radiation with average wavelength $\lambda$ has energy $h\, c / \lambda = k  T$ where
 $k$ is Boltzmann's constant and since the wavelength varies with redshift as $( 1 + z )$ this implies $T_{o} = T_{e} / (1+z)$.
   We assume that this holds even for a galaxy source which may not be a perfect blackbody. Then (\ref{eq:TobsDA_TemDL}) simplifies to
\begin{equation}
  \mathcal{D}_L = \mathcal{D}_A \, (1+z)^2  \, .    \label{eq:DL_DA_z}
\end{equation}
This relation is the same as that for FRW. Hence the angular size of a source galaxy in CGR can be found
\begin{equation} \label{eqn:ang}
\Delta{\theta} = \frac{d}{\mathcal{D}_A} = \frac{d(1+z)}{r}\left\{1-\frac{t^2}{\tau^2
}\right\}^{1/2},
\end{equation}
where (\ref{eq:lumdistance}) and (\ref{eq:DL_DA_z}) have been used.

\begin{figure}
\includegraphics[width = 5 in]{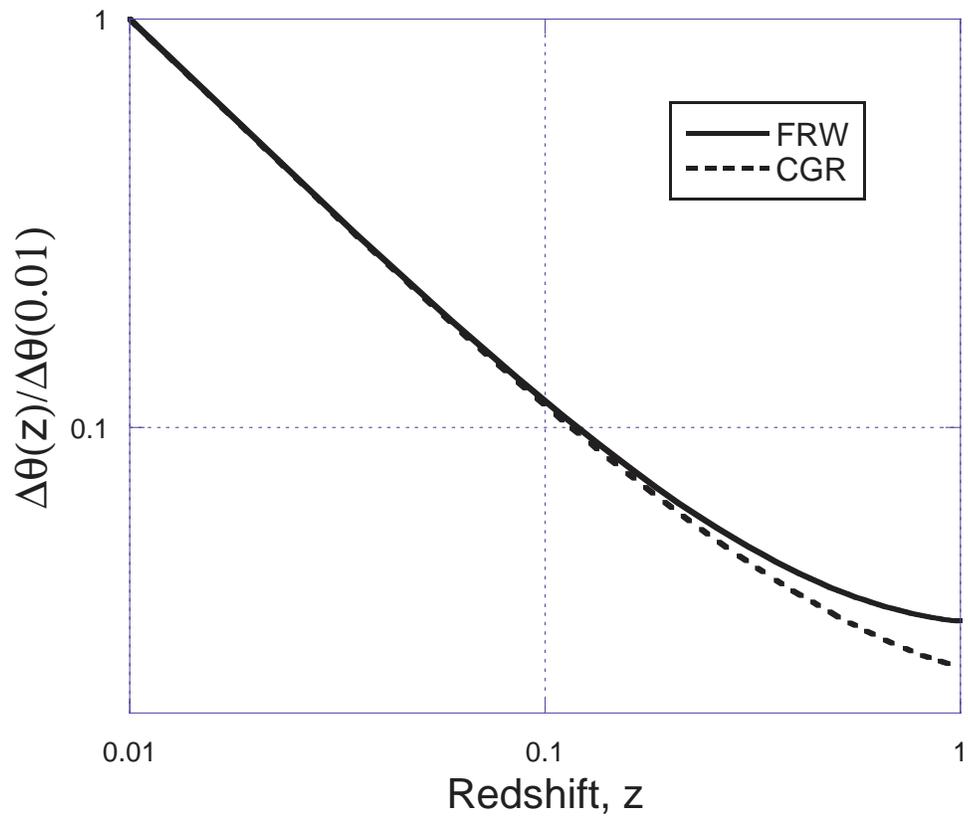}
\caption{\label{fig:fig1} Angular size shown as a function of redshift for both the FRW model (solid line) with a deceleration parameter $q_0 = 1/2$ or $\Omega_m = 1$ and the CGR model with $\Omega_{m} = 0.04$ (broken line)}
\end{figure}

Substituting (\ref{eqn:rctau2}) in (\ref{eqn:ang}) produces gravitational effects on the angular size that can be called lensing. We have plotted in Fig. \ref{fig:fig1} the dependence of angular size $\Delta \theta$ on redshift $z$ for CGR
 using (\ref{eqn:rctau2}) in (\ref{eqn:ang}) but instead with the density function $\Omega(z)$ determined by Oliveira and Hartnett  \cite{Oliveira2006}. That density expression replaces the simple form in  (\ref{eqn:rctau2}) and better characterizes the universe at high redshifts.  

In order to compare theories independently of the constants $d$, $c$ and $\tau \approx H_0^{-1}$, we plot  $\Delta \theta(z)/\Delta \theta (0.01)$ for both FRW and CGR theories. It is quite clear from Fig. \ref{fig:fig1} that for
 redshifts $z \leq 0.2$ the two models are in reasonable agreement but in general $\Delta \theta_{FRW} \neq \Delta \theta_{CGR}$.
 For $z > 0.2$ the details depend heavily on the parameters of the models chosen.
\\
\section{\label{sec:Surface}\bf{SURFACE BRIGHTNESS}}

To determine the effect of redshift variation on apparent surface brightness $B$ of a source we need to calculate the observed
 flux $\mathcal{F}_{bol}$ per unit solid angle $\Theta$,
\begin{equation} \label{eq:surfacebol_def}
  B = \frac{ \mathcal{F}_{bol} } {\Theta} \, ,
\end{equation}  
where for a source diameter of $d$ and source angular distance $\mathcal{D}_A$,
 the solid angle $\Theta$ is given by
\begin{equation} \label{eq:solidAngle}
  \Theta = \frac{\pi \, \left( d/2 \right)^2 } {\mathcal{D}^2_A} .
\end{equation}
It follows from (\ref{eq:bol}), (\ref{eq:DL_DA_z}), (\ref{eq:surfacebol_def})
 and (\ref{eq:solidAngle}) that the apparent surface brightness
\begin{eqnarray}
  B =  \frac{\mathcal{F}_{bol}}{(\pi/4)\, d^2 / \mathcal{D}^2_A}  
    = \frac{L_{bol}} {\pi^2 d^2} \, \left( 1 + z \right)^{-4} \,  ,\label{eq:surfacebol} 
  \end{eqnarray}
which is the same as the usual FRW expression, the same $(1+z)^{-4}$ dependence Tolman \cite{Tolman1930} produces using standard 
cosmology.  
\\

\section{\bf{DENSITY}}

  In terms of the phase space expansion history, the universe at time $t$ 
  has a total relativistic mass $M$ and a total volume $V$. The expansion is assumed to be 
  symmetric so that the volume $V$ is spherical.
  The average matter density $\rho$ is
  \begin{equation}
     \rho = \frac{M}{V} \, .  \label{eq:rho=M/V}
  \end{equation}

  The total relativistic mass of matter $M$ in Cosmological Special Relativity \cite{Carmeli2002}
  at cosmic time $t$ is 
  \begin{equation}
   M = \frac{M_{0}}{\sqrt{1-t^2 / \tau^2 \,}} \, , \label{eq:M}
  \end{equation} 
  where $M_{0}$  is the mass of the universe at the present epoch $t=0$.

  The volume is taken to be that of a sphere
  \begin{equation}
     V = \frac{4 \pi}{3\,} R^{3} \, , \label{eq:volume}
  \end{equation}
  where $R$ is the radius of the portion of the universe that just
  contains the mass $M$.  In CGR, the distance $r$ is measured from the observer at
  the present epoch to the source rather than the other way, e.g. as is done in the Friedmann theory
  of cosmology. We assume that higher density corresponds to higher velocity
  and that the volume decreases as velocity increases. The radius $R$ of
  the universe is therefore taken to be
  \begin{equation}
     R = c\, \tau  - r \, , \label{eq:R}
  \end{equation}
  where the redshift distance relationship $r$ is given by (\ref{eqn:rctau}).

  $R$ is defined this way so that for $v = 0$, $R(r=0) = c\, \tau$
  is the radius of the sphere of the universe that just contains the mass of
  matter $M_{0}$. We define the average matter density parameter
  \begin{equation}
    \Omega = \frac{\rho}{\rho_{c}} \, ,   \label{eq:Omega=rho/rho_c}
  \end{equation}
  where $\rho_{c} = 3/\left( 8 \pi\, {\rm G\,} \tau^2 \right)$ is the critical density.
  An overall constraint is that, for $\Omega \geq 0$,
  \begin{equation}
    1 + \frac{\left( 1 - \Omega \right)}{c^{2} \tau^{2}} r^{2}\,  > \, 0 \, .
         \label{eq:1+f(r,Omega<1)>0}  
  \end{equation}
 
  From (\ref{eq:rho=M/V})-(\ref{eq:Omega=rho/rho_c}) the function for $\Omega$ is
  \begin{equation}
     \Omega = \frac{\Omega_{m} / \sqrt{1-\beta^{2}\,}}
      {\left[ 1 - {\rm sinh}\left(\beta \sqrt{1 - \Omega}\,\right)
                / \sqrt{1 - \Omega}\, \right]^{3} } \, ,
             \label{eq:Omega_Omega_m_expanded}
  \end{equation}
  where
  \begin{eqnarray}
    \Omega_{m} &=& \frac{\rho_{m}}{\rho_{c}}  \, , \label{eq:Omega_m}  \\
  \nonumber \\
    \rho_{m} &=& \frac{M_{0}}{\left(4 \pi / 3\right) \left( c\, \tau\right)^{3}} \, ,
             \label{eq:rho_m}
  \end{eqnarray}
  where $\rho_{m}$ is the average matter density at the current epoch.

  In the first order approximation where $\beta \ll 1$, $z \approx \beta$.
  Since ${\rm sinh}(x) \approx x$ for small $x$,
  (\ref{eq:Omega_Omega_m_expanded}) can be written
  \begin{eqnarray}
    \Omega \approx \frac{\Omega_{m} \left( 1 + (1/2)\,\beta^{2} \right)}
           {\left( 1 - \beta \right)^{3}}
           \approx \Omega_{m} \left( 1 + z \right)^{3} \, .
  \end{eqnarray}
  In the Friedmann-Robertson-Walker cosmologies, the matter density parameter
  $\Omega = \Omega_{m} (1 + z)^3$ for all z in a dust dominated spatially flat universe, 
  but this is not the case in the present
  theory where the density varies more strongly than $(1+z)^3$. This will produce 
  significant results in the data analysis.

  The derived relation (\ref{eq:Omega_Omega_m_expanded}) for $\Omega$ is transcendental. For fits
  to data it is more convenient to have a regular function, hence we use a second
  order approximation for $\Omega$, which is briefly described in appendix (\ref{ap:Omega_approx}).
 
\section{\label{sc:Redshift_z_t} \bf{EXPANSION TRANSITION REDSHIFT $z_t$}}

  In CGR the expansion has three basic phases: decelerating, coasting and
  finally accelerating, corresponding to density $\Omega > 1$, $\Omega = 1$, and $\Omega < 1$,
  respectively \cite{carmeli-5}.  What is the expected velocity and redshift of the transition
  from deceleration to acceleration?  This phase shift occurs during the zero acceleration or coasting
  phase when $\Omega = 1$. Taking (\ref{eq:Omega_Omega_m_expanded}) to the limit
  $\Omega \rightarrow 1$, since ${\rm sinh}(x) \approx x$ for small $x$, yields
  \begin{equation}
    \lim_{\Omega \rightarrow 1}{\Omega} = 1
                       =  \frac{\Omega_{m} / \sqrt{1-\beta_{t}^{2}\,}}
                               {\left( 1 - \beta_{t}\right)^{3}} \, , \label{eq:Omega=1}
  \end{equation}
  which simplifies to
  \begin{equation}
     \left( 1 - \beta_{t} \right)^{3} \, \sqrt{ 1 - \beta_{t}^{2}\,} =  \Omega_{m} \, .
          \label{eq:beta_at_Omega=1}
  \end{equation}
  Solving (\ref{eq:beta_at_Omega=1}) for $\beta_{t}$, the predicted redshift $z_{t}$
  of the expansion transition is obtained from  (\ref{eqn:defbeta}).

\section{\label{sec:Comparison} \bf{COMPARISON WITH HIGH-Z TYPE Ia SUPERNOVAE DATA}}

The redshift distance relationship in CGR is given by (\ref{eqn:rctau}) and $\Omega$ is evaluated from (\ref{eq:Omega_2_ex}). 

In order to compare the redshift distance relation with the high redshift SNe Ia data from Riess \textit{et al} \cite{Riess2004}
 and Astier \textit{et al} \cite{Astier2005}, the proper distance is converted
 to magnitude as follows.
\begin{equation} \label{eqn:lumindistance}
m(z) = \mathcal{M} + 5\,log \left[ \mathcal{D}_{L}(z;\Omega) \right], 
\end{equation}
where $\mathcal{D}_{L}$ is the dimensionless ``Hubble constant free'' luminosity distance (\ref{eq:lumdistance}).
 Refer \cite{Perlmutter1997,Riess1998}.
Here 
\begin{equation} \label{eqn:scriptem}
\mathcal{M} = 5\,log(\frac{c \tau}{Mpc}) + 25 + M_{B} + a.
\end{equation}
The units of $c \tau$ are $Mpc$. The constant $25$ results from the luminosity distance expressed in $Mpc$. However,
 $\mathcal{M}$ in (\ref{eqn:lumindistance}) represents a scale offset for the distance modulus (m-$M_{B}$). It is 
sufficient to treat it as a single constant chosen from the fit. In practice we use $a$, a small free parameter, to
 optimize the fits. From (\ref{eq:lumdistance}), with $\beta = t/\tau$ the luminosity distance is given by

\begin{equation} \label{eqn:magnitude}
\mathcal{D}_{L}(z;\Omega_{m}) = \frac {r} {c \tau} (1+z)\left(1-\beta^2 \right)^{-1/2}  
\end{equation}
using (\ref{eqn:rctau}), hence $r$ in units of $c \tau$. $\mathcal{D}_{L}$ is only a function of $\Omega_m$ and $z$.

The parameter $\mathcal{M}$ incorporates the various parameters that are independent of the redshift, $z$. The parameter
 $M_{B}$ is the absolute magnitude of the supernova at the peak of its light-curve and the parameter $a$ allows for any
 uncompensated extinction or offset in the mean of absolute magnitudes or an arbitrary zero point. The absolute magnitude
 then acts as a ``standard candle'' from which the luminosity and hence distance can be estimated. 

The value of $M_{B}$ need not be known, neither any other component in $\mathcal{M}$, as $\mathcal{M}$ has the effect of
 merely shifting the fit curve (\ref{eqn:magnitude}) along the magnitude axis. 

However by choosing the value of the Hubble-Carmeli constant $\tau = 4.28 \times 10^{17} \, s = 13.58 \; Gyr$, which is
 the reciprocal of the chosen value of the Hubble constant in the gravity free limit $h = 72.17 \pm 0.84$ (statistical)
 $km.s^{-1} Mpc^{-1}$ (see Section \ref{sec:Hubbleconst}) $\mathcal{M} = 43.09 + M_{B} + a$.  

We use two SNe Ia data sets for the curved fitting analysis. The data are drawn from Table 5 of Riess \textit{et al}  \cite{Riess2004}, the Supernova Cosmology Project, and Tables 8 and  9 of Astier \textit{et al} \cite{Astier2005}, the Supernova Legacy Survey (SNLS). Also we combined the data  sets of Riess \textit{et al} and Astier \textit{et al} and found the best statistical fit to all those data.

\begin{figure}
\includegraphics[width = 5 in]{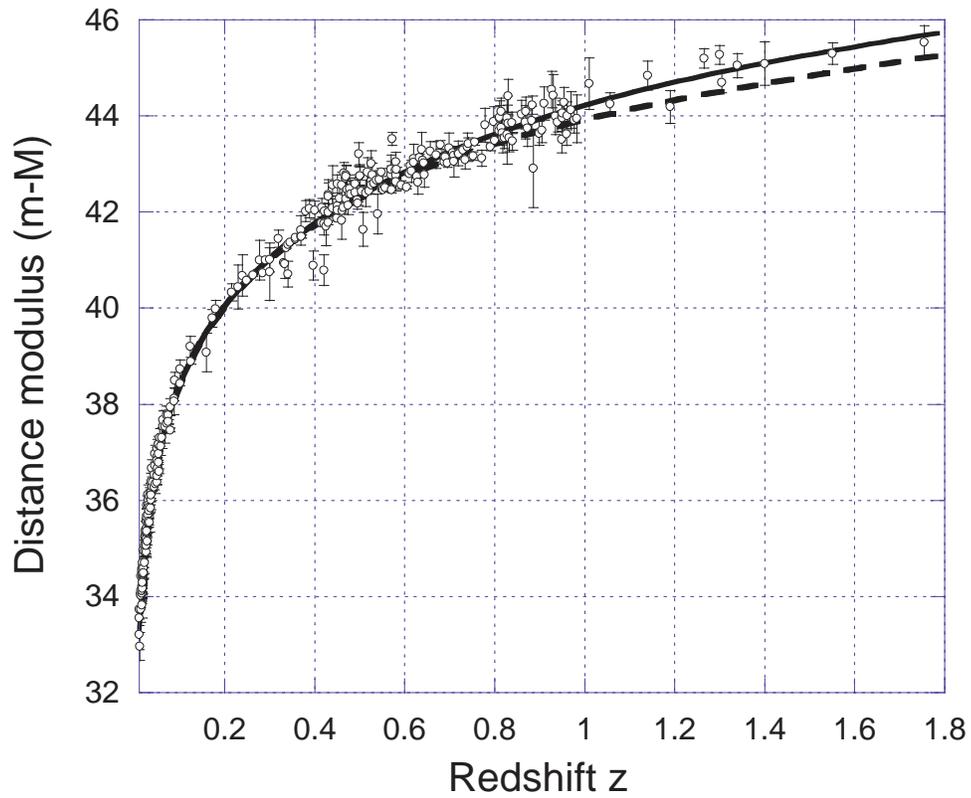}
\caption{\label{fig:fig5} The combined data sets of Riess \textit{et al} and Astier \textit{et al}.  The solid line represents the statistically best fit curve with $a=0.2284$ and  $\Omega_m =0.0401$ and the broken line represents the curve with $a=0.2284$ and $\Omega_m =0.263$}
\end{figure}

\begin{figure}
\includegraphics[width = 5 in]{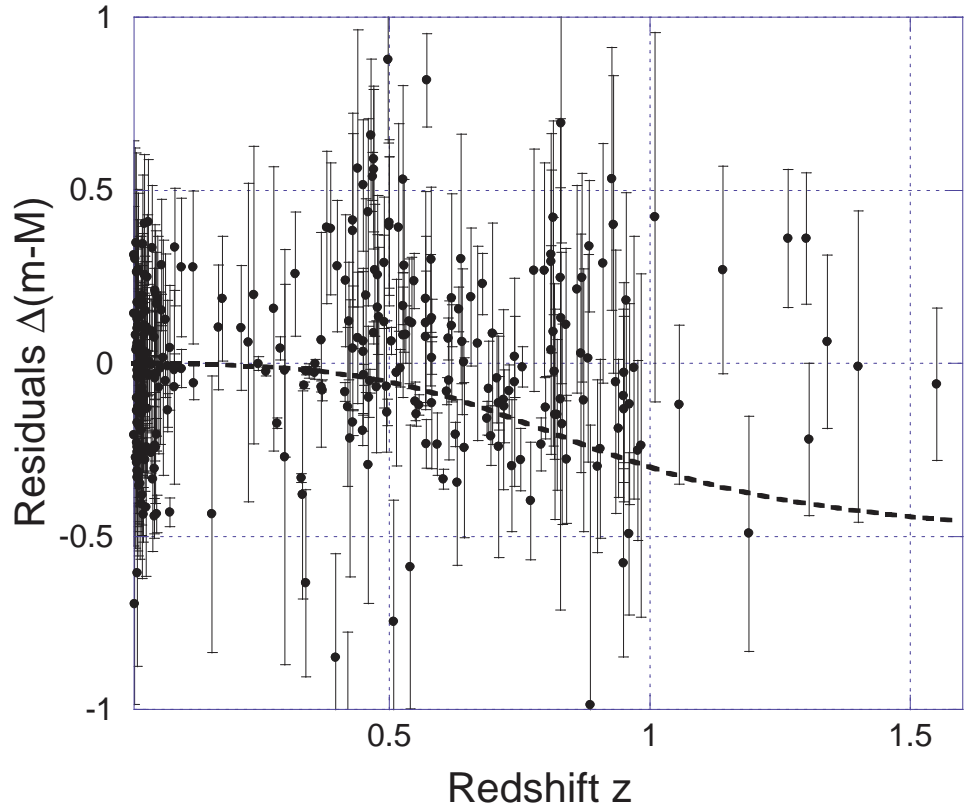}
\caption{\label{fig:fig6} Residuals vs redshift (on linear scale): the differences between the best fit curve with $\Omega_m = 0.0401$ and $a=0.2284$ and the data of fig. \ref{fig:fig5}. The mean of the residuals is $8.04 \times 10^{-5}$ when all errors are assumed equal and $-0.0769$ when weighted by published errors. The broken line represents the curve where $\Omega_m =0.263$}
\end{figure}

 This is shown in fig. \ref{fig:fig5} along with the curve where $\Omega_m = 0.263$, which is the value that Astier \textit{et al} quote for the average matter density at the current epoch. Lastly, we take the residuals between the combined the data set of Riess \textit{et al} and Astier \textit{et al} and the best fit curve of fig. \ref{fig:fig5}. This is shown in fig. \ref{fig:fig6}, along with the curve that represents $\Omega_m = 0.263$.

\section{\label{sec:Fits}\bf{QUALITY OF CURVE FITS}}

In order quantify the goodness of the least squares fitting we have used the $\chi^{2}$ parameter which measures the goodness of the fit between the data and the theoretical curve assuming the two fit parameters $a$ and $\Omega_m$. Hence $\chi^{2}$ is calculated from
\begin{equation} \label{eqn:chisq}
\chi^2 = \sum_{i=1}^{N} \frac{1}{\sigma_i^2} \left[(m-M)(z)_i-(m-M)(z_{obs})_i\right]^2,
\end{equation}
where $N$ are the number of data; $(m-M)(z)$ are determined from (\ref{eqn:lumindistance}) with fit values of $a$ and $\Omega_m$;
 $(m-M)(z_{obs})$ are the observed distance modulus data at measured redshifts $z_{obs}$; $\sigma_i$ are the published magnitude 
errors. The values of $\chi^2/N$ ($\approx \chi_{d.o.f}^2$) are shown in Table I, calculated using published errors on the distance modulus data. In each case the best fit value of $a$ is found for each value of $\Omega_m$.

Table I lists the $\chi^2/N$ parameters determined for three values of $\Omega_m$, as well as the best fit values of $\Omega_m$ determined using the Mathematica software package. The latter are indicated by the word `best' in the table.  In the latter case  the best fits are only statistically determined and hence also the standard error. In all instances the best fit value was determined  for the parameter $a$. 

From the combined data set of  Riess \textit{et al} and Astier \textit{et al} the best statistical fit resulted in a value of $\Omega_m = 0.0401 \pm 0.0199$, which is consistent with the result obtained by averaging the values of $\Omega_m$ obtained from the individual data sets. 

\begin{table}[ph]
\center
\small
Table I:~ \hspace{4pt} Curve fit parameters\\
\vspace{6pt}
\begin{tabular}{c|c|cccccc} \hline \hline
Data set		&N					& a				&$\sigma(stat)$& $\Omega_m$& $\sigma(stat)$&	$\chi^2$/N	&$\chi^2$/N ($\sigma_i =1$)\\
\hline \hline
Riess \textit{et al}
						&185				& 0.257 	&							& 0.021 	& 							&1.34188 \\
						&						& 0.268		&							& 0.042 	& 							&1.32523 \\
best				&						& 0.278 	&	0.025				& 0.0631 	& 0.0303 				&1.32152 \\
\hline
Astier \textit{et al}						
						&117				& 0.158 	&							& 0.021 	& 							&11.2656 \\
best				&						& 0.161 	&	0.043				& 0.0279 	& 0.0430				&11.3199 \\
						&						& 0.168		&							& 0.042 	& 							&11.4533 \\
						&						& 0.177		&							& 0.063 	& 							&11.6919 \\
\hline
Riess + Astier
						&302				& 0.219 	& 						& 0.021 	& 							&6.70338 &0.075039\\
best				&						& 0.228		& 0.018				& 0.0401 	& 0.0199				&6.99446 &0.074726\\
						&						& 0.229 	& 						& 0.042 	& 							&7.02192 &0.074728\\
						&						& 0.239		& 						& 0.063 	& 							&7.32371 &0.075010\\
						&						& 0.304		& 						& 0.263 	& 							&10.0568 &0.086165\\
\hline

\end{tabular}
\end{table}

The differences in the relative magnitudes of the $\chi^2$/N values for each data set is primarily the result of the size of the  published errors used in the calculation (\ref{eqn:chisq}) in the Astier \textit{et al} data set. The published errors for Astier \textit{et al} data are small in relation to their deviation from the fitted curve, as evidenced by their large $\chi^2$/N values compared to Riess \textit{et al} data in Table I. Hence it appears that Astier \textit{et al} have underestimated the real errors in their data.

Looking at the $\chi^2$/N values the minimum regions in each set overlap where $\Omega_m = 0.042$. This is then the region of  the most probable value. This is consistent with a value of $\Omega_m = 0.0401 \pm 0.0199$ as determined from the combined data set shown in fig. \ref{fig:fig5}. Therefore no exotic dark matter need be assumed as this value is within the limits of the locally 
measured baryonic matter budget $0.007 < \Omega_m < 0.041$ \cite{Fukugita1998} where a Hubble constant of $70 \; km.s^{-1} Mpc^{-1}$  was assumed.

Previously one of us \cite{Hartnett2006}, which used some of the same data but with a different density model, the $\chi^2$/N parameters appear to be much smaller and therefore represent better quality fits than in the former. However this is not actually the case, as a software algorithm was used in \cite{Hartnett2006} that didn't properly calculate $\chi^2$. The problem with the analysis was that the errors for all data were set to unity, that is, $\sigma_i = 1$. In Oliveira and Hartnett \cite{Oliveira2006} we calculated the correct $\chi^2/N$ parameters using (\ref{eqn:chisq}) and published errors. 

So for a comparison here, the $\chi^2/N$ parameters, where $\sigma_i$ are forced to unity, are also shown in Table I. The resulting $\chi^2/N (\sigma_i=1)$ are extremely good even compared to the 185 data of Riess \textit{et al} fitted to in Fig. 1 of Hartnett \cite{Hartnett2006} where $\chi^2/N (\sigma_i=1) = 0.2036$ was calculated. 

The improvement has resulted from the additional factor $(1-t^2/\tau^2)^{-1/2}$ in the luminosity distance and a little from the refinement of the density model $\Omega(z)$. If we exclude the new density model and use $\Omega=\Omega_m(1+z)^3$ where $\Omega_m = 0.04$ instead, we get $\chi^2/N (\sigma_i=1) = 0.075986$ for the best fit to the combined data set requiring $a = 0.2152$. This indicates the improvement over Hartnett \cite{Hartnett2006} is more the result of the additional factor in the luminosity distance than the better density model.

Looking at the curve fits of fig. \ref{fig:fig5} where the distance modulus vs redshift curves with both $\Omega_m = 0.0401$ and  $\Omega_m = 0.263$ are shown, it is quite clear that using the Carmeli theory a universe with $\Omega_m = 0.263$ is ruled out and
 hence also the need for any dark matter.  This is even more obvious from the residuals shown in fig. \ref{fig:fig6}. There the fit
 with $\Omega_m = 0.0401$ is drawn along the $\Delta(m-M)=0$ axis and the fit with $\Omega_m = 0.263$ is shown as a broken line. The
 highest redshift data clearly rules out such high matter density in the universe.

The best fit result of this paper, $\Omega_m = 0.0401 \pm 0.0199$, with a density function that is valid for all $z$ over the range of observations, is also consistent with the result obtained by Hartnett \cite{Hartnett2006} $\Omega_m = 0.021 \pm 0.042$ but here
 the 1 $\sigma$ errors are significantly reduced. 

 With the best fit $\Omega_{m} = 0.0401$, the predicted expansion transition redshift from (\ref{eq:beta_at_Omega=1}) is
 \begin{equation}
   z_{t} = 1.095\, {}^{+0.264}_{-0.155} \, .
 \end{equation}
 This is about a factor of $2$ greater than the fitted value reported by Riess \textit{et al.}\cite{Riess2004} of $z_{t} = 0.46 \pm 0.13$,
 which was from a best fit to the differenced distance modulus data, a second order effect.
 They used a luminosity distance relation assuming a flat Euclidean space (i.e., $\Omega_{total}=1$) and fit
 the difference data with the deceleration parameter $q(z) = (dH^{-1}(z)/dt) - 1$ .

  In the present theory, the transition redshift $z_{t}$ is inherently where the density parameter $\Omega(z_t) = 1$.   Thus, the transition
 is determined simultaneously with the initial fit of $\mathcal{D}_{L}$ to the data.

Moreover $\Omega_m$ has been determined as a `Hubble constant free' parameter because it comes from $\mathcal{D}_{L}(z;\Omega_{m})$, which
 is evaluated from fits using (\ref{eqn:magnitude}). The latter is independent of the Hubble constant or more precisely in this theory
 $\tau$ the Hubble-Carmeli time constant. Therefore $\Omega_m$ should be compared with $\Omega_b$ and not with $\Omega_b h^2$, where $h$
 is the Hubble constant as a fraction of $100 \; km.s^{-1} Mpc^{-1}$ and not to be confused with $h = 1/\tau$ used in CGR.

Nevertheless the value of $\Omega_b h^2 = 0.024$ from \cite{Spergel2006} and $h =0.7217$ (assuming a value of
 $\tau^{-1} = 72.17 \; km.s^{-1}Mpc^{-1}$) implies $\Omega_b = 0.043$, which is in good agreement with the results of this work.
 Yet caution must be advised as the problem of the analysis of the WMAP data has not yet been attempted within the framework of CGR.

\section{\bf{VALUES OF SOME KEY PARAMETERS}}

\subsection{\label{sec:Hubbleconst}Hubble constant}

Using the small redshift limit of (\ref{eqn:rctau}) and the Hubble law at small redshift ($v = H_{0}r$) it has been shown
 \cite{carmeli-6} that the Hubble parameter $H_0$ varies with redshift. If it applies at the low redshift limit it follows
 from the theory that at high redshift we can write
\begin{eqnarray} \label{eqn:H0-h}
H_{0}= h \frac {\beta \sqrt{1-\Omega}}{\sinh (\beta \sqrt{1-\Omega})}.
\end{eqnarray}
Therefore $H_{0}$ in this model is redshift dependent, not constant and $H_{0} \leq h$. Only $h = \tau^{-1}$ is truly independent of
 redshift and constant. The condition where $H_{0} = h$ only occurs at $z = 0$ and where $\Omega \rightarrow 0$.

\begin{figure}
\includegraphics[width = 5 in]{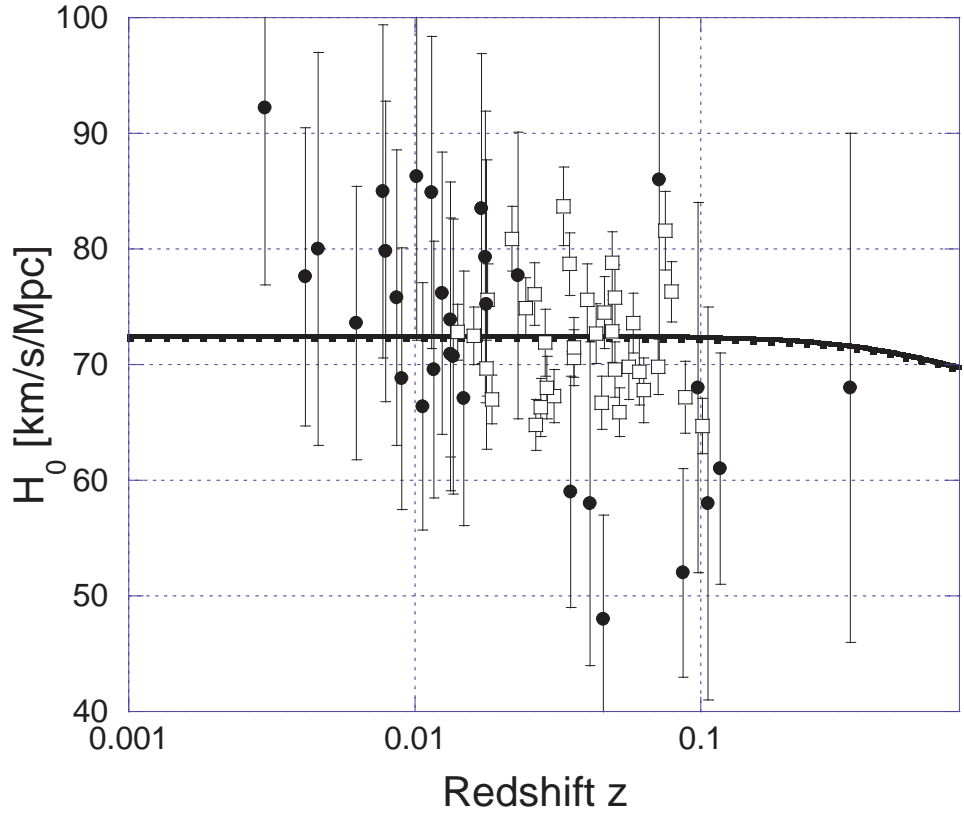}
\caption{\label{fig:fig7} Hubble constant $H_{0}$ as a function of redshift, $z$. The filled circles are determined from Tully-Fisher measurements taken from \cite{Freedman1994}, Table 5 of \cite{Tutui2001} and Table 7 of \cite{Freedman2001}, except the point at z = 0.333 is from Sunyaev-Zel'dovich effect taken from Fig. 4 of \cite{Tutui2001}. The open squares are determined from the SN Ia measurements and taken from Table 6 of \cite{Freedman2001} and Table 5 of \cite{Riess2004}. The errors are those quoted in the sources from which the data was taken}
\end{figure}

By plotting $H_{0}$ values determined as a function of redshift, using (\ref{eqn:H0-h}), it is possible to get an independent determination
 of $h$, albeit the noise in the data is very large. This is shown in fig. \ref{fig:fig7} with values calculated by two methods with the
 exception of one point at $z = 0.333$. See figure caption for details. The data, even though very scattered, do indicate a trending down 
of $H_{0}$ with redshift.

Separate curve fits from (\ref{eqn:H0-h}), with $h$ as a free parameter, have been applied to the two data sets, Tully-Fisher (TF) (the 
solid line)  and SNe type Ia (the broken line) measurements. The former resulted in $h = 72.47 \pm 1.95$ (statistical) $\pm 13.24$ (rms)
 $km.s^{-1}Mpc^{-1}$ and from the latter $h = 72.17 \pm 0.84$ (statistical) $\pm 1.64$ (rms) $km.s^{-1}Mpc^{-1}$. The rms errors are those 
derived from the published errors, the statistical errors are those due to the fit to the data alone. The SNe Ia determined value is more
 tightly constrained but falls within the TF determined value.

\subsection{Mass of the universe}

  It is easily shown from (\ref{eq:Omega_m}) and (\ref{eq:rho_m}) that
  \begin{equation}
  \Omega_m = R_{s}/R_{0} \, , \label{eq:Omega_m=Rs/R0}
  \end{equation}
  where $R_{s}=2\,G\,M_{0}/c^2$ is
  the Schwarzschild radius if the present universe rest mass $M_{0}$ is imagined to
  be concentrated at a point, and $R_{0} = c\, \tau$ is the present radius of the
  universe.  From this we get the present universe rest mass
  \begin{equation}
   M_{0} = \Omega_{m} \frac{c^{3} \tau}{2\, G} \, , \label{eq:M_0}
  \end{equation}
  which, with $\Omega_{m} = 0.0401 \pm 0.0199$ gives
  \begin{equation}
   M_{0} = \left(  1.74 \pm 0.86 \right) \times 10^{21} M_{\odot} \, . \label{eq:M_0_value}
  \end{equation}
  Likewise, the average matter density (\ref{eq:Omega_m})
  \begin{equation}
    \rho_{m} = \Omega_{m} \rho_{c} = \left(3.92 \pm 1.94 \right) \times 10^{-31} {\rm gm}\, {\rm cm}^{-3} \, .
  \end{equation}

\subsection{Time of transition from deceleration to acceleration}

  From Carmeli's cosmological special relativity \cite{Carmeli2005} we get a relation
  for the cosmic time in terms of the redshift.  In particular, in terms of $z_{t}$ we have for
  the cosmic time $t_{t}$ of the expansion transition from the present
  \begin{equation}
    t_{t} = \tau\, \frac{\left( 1 + z_{t} \right)^2 - 1}{\left( 1 + z_{t} \right)^2 + 1}
                 \, .\label{eq:t_t}
  \end{equation}
  For the above value of $z_{t}$ and for the age of the universe $\tau = 13.58\, {\rm Gyr}$ we have
  \begin{equation}
    t_{t} =  8.54^{+0.903}_{-0.662}\, {\rm Gyr} \, .
  \end{equation}
  Since the big bang ($t^{*}=0$), the transition cosmic time is $t^{*}_{t} = \tau - t_{t}$,
  \begin{equation}
    t^{*}_{t} = 5.04^{+0.662}_{-0.903}\, {\rm Gyr} \, .
  \end{equation}

 In Fig. \ref{fig:fig8} is a plot of the density for $\Omega_{m}=0.04$.
 More than $8.54\, {\rm Gyr}$ ago the density was higher than the critical value ($\Omega > 1$ .)
 Since the transition the density has become less than critical ($\Omega < 1$). The fit to the
 SNe Ia data was accomplished without the need for any dark energy, usually associated with
 the cosmological constant. In CGR there is no cosmological constant although a value for it may be
 obtained by a comparison study \cite{carmeli-4, Hartnett2006}.

\section{\label{sec:conclusion}\bf{CONCLUSION}}

 The surface brightness is the same as in standard cosmology, though angular size is smaller by
 a factor of $(1-t^2/\tau^2)^{1/2}$.

The analysis in this paper has shown that the most probable value of the local density of the Universe is $\Omega_m = 0.0401 \pm 0.0199$ the best fit from a combined data set of two totaling 302 data. The fits used a density function with
 limited range and validity and did not take into account the published errors on the individual magnitude data.  The fits to the data
 are consistent over the entire range of the available redshift data, from $0.1 < z < 2.0$, a result of the more accurate relation
 for $\Omega$, as well as the proper accounting of the increase in the source luminosity due to the factor $\left(1-\beta^2\right)^{-1/2}$.

Since $\Omega_{m}$ is within the baryonic matter density budget, there is no need for any dark matter to account for the SNe Ia redshift
 magnitude data. 
Furthermore, since the predicted transition redshift $z_{t}  = 1.095\, {}^{+0.264}_{-0.155}$ is well within the redshift range of the data,
 the expansion rate evolution from deceleration to acceleration, which occurred  about $8.54\, {\rm Gyr}$ ago, is explained
without the need for any dark energy.

 The density $\Omega_{m} < 1$ and the determination of the transition redshift $z_{t}$ within the data support the conclusion
 that the expansion is now accelerating and that the universe is, and will remain open.

\appendix

\section{\bf{APPROXIMATION OF $\Omega$} \label{ap:Omega_approx}}

\begin{figure}
\includegraphics[width = 5 in]{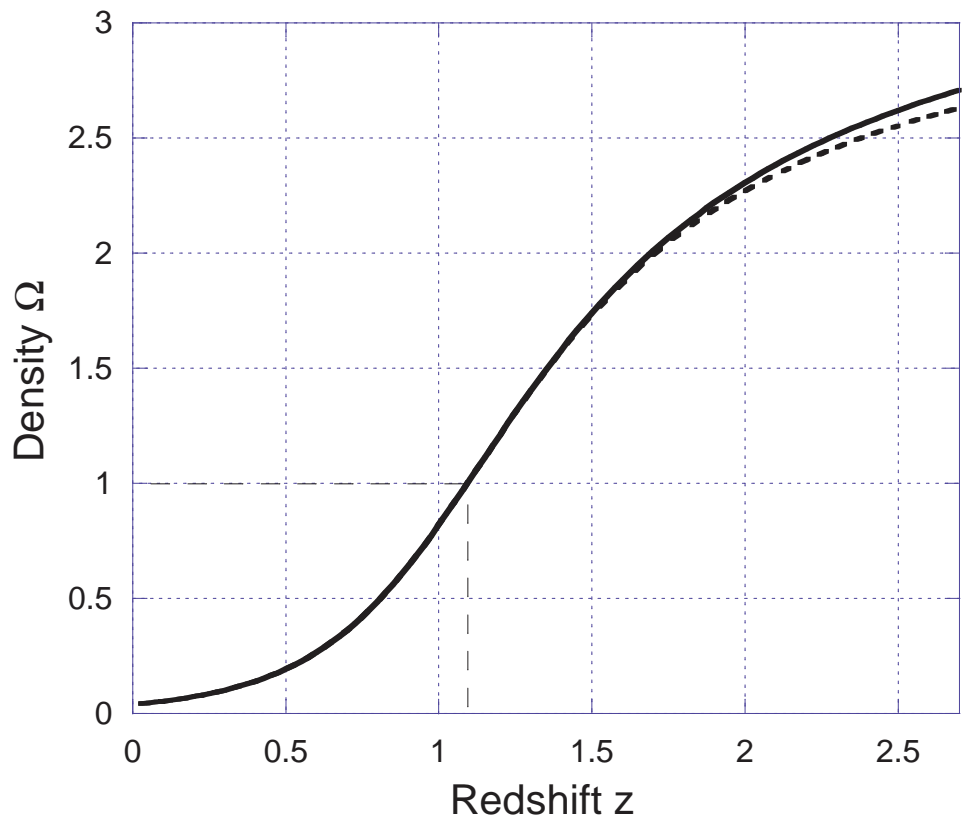}
\caption{\label{fig:fig8} Density model shown as function of redshift for both approximated (broken line) and exact (solid line) with the same value of $\Omega_{m} = 0.04$. The transition redshift $z_t = 1.095$ where $\Omega = 1$ is indicated by the dashed lines}
\end{figure}

 The form for $\Omega$ in (\ref{eq:Omega_Omega_m_expanded}) is transcendental,
 which is not convenient for fitting. A second order approximation can be made
 by taking ${\rm sinh}(x) \approx x + x^{3}/3!\,$. With this approximation
 (\ref{eq:Omega_Omega_m_expanded}) becomes
 \begin{eqnarray}
    \Omega &\approx& \Omega_{2} = \frac{\Omega_{m} / \sqrt{1 - \beta^{2}\,}}
       {\left\{1 - \left[ \beta \sqrt{1 - \Omega_{2}}
           + \beta^{3} \left( \sqrt{1 - \Omega_{2}}\, \right)^{3} / 3! \right]
                / \sqrt{1 - \Omega_{2}}\,  \right\}^{3}} \, , \label{eq:Omega_2}
 \end{eqnarray}
 which simplifies to
 \begin{eqnarray}
   \Omega_{2} \left[ 1 - \beta
    - \frac{\beta^{3}}{ 3!} +  \frac{\beta^{3}}{ 3!} \Omega_{2} \right]^{3}
           - \left( \Omega_{m} / \sqrt{1 - \beta^{2}\,} \right) = 0 \, . \label{eq:Omega_2_ex}
 \end{eqnarray}
 This is a quartic equation in $\Omega_{2}$ and can be solved for $\Omega_{2}$ as a function
 of $\beta$ by standard methods. $\Omega_{2}$ is shown in fig. \ref{fig:fig8} as the broken line
 where a matter density $\Omega_m = 0.04$ was assumed. It is compared with $\Omega$ given by
 the exact form (\ref{eq:Omega_Omega_m_expanded}).

\newpage

\end{document}